\documentclass[12pt]{article}
\usepackage{graphics}
\usepackage{amsmath}

 \font\sevenrm=cmr7
\newif\ifinexp \inexpfalse
\newcommand\I{\ifinexp \hbox{\hskip0.8pt\sevenrm i} \else
\hbox{\hskip1pt\rm i} \fi}
\newcommand\E[1]{\inexptrue \hbox{e}^{#1} \inexpfalse}
\newcommand\D{\displaystyle}
\newcommand\hr{\hbox{\textbf{R}}}

\newtheorem{lemma}{\bf Lemma}
\newtheorem{theorem}{\bf Theorem}
\newtheorem{remark}{\bf Remark}
\newenvironment{demo}{\textbf{Proof. }}{\vskip4pt}

\title{Darboux transformations and global solutions for
a nonlocal derivative nonlinear Schr\"odinger equation}

\author{\small Zi-Xiang Zhou\\
\small School of Mathematical Sciences, Fudan University, Shanghai
200433, China\\\small Email: zxzhou@fudan.edu.cn}

\date{}

\begin{document}

\maketitle

\begin{abstract}
A nonlocal derivative nonlinear Schr\"o\-dinger equation is
introduced. By constructing its basic Darboux transformations of
degrees one and two, the explicit expressions of new solutions are
derived from seed solutions by Darboux transformation of degree
$2n$. Usually the derived solutions of this nonlocal equation may
have singularities. However, by suitable choice of eigenvalues and
the parameters describing the ratio of the two entries of the
solutions of the Lax pair, global bounded solutions of the nonlocal
derivative nonlinear Schr\"odinger equation are obtained from zero
seed solution by a Darboux transformation of degree $2n$.
\end{abstract}

\section{Introduction}

In \cite{bib:AbMu}, Ablowitz and Musslimani introduced the nonlocal
nonlinear Schr\"o\-din\-ger equation and got its explicit solutions
by inverse scattering. Quite a lot of work were done after
that.\cite{bib:AbMu2,bib:HuangLing,bib:Khare,bib:LiXu,bib:Sarma}
Various other nonlocal integrable equations are also
considered,\cite{bib:AbMu2,bib:Gadz} including high dimensional
equations\cite{bib:Fokashigh} and discrete
equations.\cite{bib:Khare,bib:Zhu}

The derivative nonlinear Schr\"odinger
equation\cite{bib:KN,bib:Mjolhus}
\begin{equation}
   \I q_t(x,t)=q_{xx}(x,t)+\I\varepsilon(q(x,t)^2q^*(x,t))_x
   \quad(\varepsilon=\pm 1)
   \label{eq:DNLS0}
\end{equation}
is an important integrable equation which describes the Alfv\'en
wave in plasma physics. In this paper, we propose an integrable
equation
--- a nonlocal derivative nonlinear Schr\"odin\-ger equation
\begin{equation}
   \D\I q_t(x,t)=q_{xx}(x,t)+\varepsilon(q(x,t)^2q^*(-x,t))_x
   \quad(\varepsilon=\pm 1).
   \label{eq:DNLS}
\end{equation}
Here ${}^*$ means complex conjugation. Although this equation is not
invariant under $q(x,t)\to q^*(-x,-t)$, it has a conserved density
$q(x,t)q^*(-x,t)$, which is invariant under spacial reversion
together with complex conjugation as that for the nonlocal nonlinear
Schr\"odinger equation. Note that the coefficient of the nonlinear
term in (\ref{eq:DNLS}) is real, rather than purely imaginary in the
usual derivative nonlinear Schr\"odinger equation (\ref{eq:DNLS0}).

In Section~\ref{sect:LP} of this paper, the Lax pair for the
nonlocal derivative nonlinear Schr\"odinger equation is presented
and its symmetries are considered. In Section~\ref{sect:1},
Section~\ref{sect:2} and Section~\ref{sect:2n}, the Darboux
transformations of degree one, two and $2n$ are discussed
respectively and explicit expressions of the new solutions are
derived. In general, the derived solutions may have singularities.
In Section~\ref{sect:eg}, some global solutions are obtained from
zero seed solution by Darboux transformations of degree two, four
and eight respectively with suitable choice of parameters. In
Section~\ref{sect:global}, we prove that the solutions given by
Darboux transformations of arbitrary degree $2n$ from zero seed
solution are globally defined and bounded for $(x,t)\in\hr^2$ if the
arguments of all eigenvalues are $\pi/4$ and the parameters
describing the ratio of the two entries of the solutions of the Lax
pair are small enough.

\section{Lax pair and its symmetries}\label{sect:LP}

Consider the Lax pair
\begin{equation}
   \Phi_x=U\Phi,\quad \Phi_t=V\Phi \label{eq:LP}
\end{equation}
where
\begin{equation}
   \begin{array}{l}
   \D U=\lambda^2J+\lambda P=\left(\begin{array}{cc}\lambda^2&\lambda q\\
   \lambda r&-\lambda^2\end{array}\right),\\
   \D V=\left(\begin{array}{cc}
   -2\I\lambda^4+\I qr\lambda^2&-2\I q\lambda^3+(-\I q_x+\I q^2r)\lambda\\
   -2\I r\lambda^3+(\I r_x+\I qr^2)\lambda&2\I\lambda^4-\I qr\lambda^2
   \end{array}\right),
   \end{array}\label{eq:LPcoeff}
\end{equation}
$q,r$ are functions of $(x,t)$, and $\lambda$ is a spectral
parameter.

The integrability condition $U_t-V_x+[U,V]=0$ gives the evolution
equations
\begin{equation}
   \begin{array}{l}
   \D\I q_t=q_{xx}-(q^2r)_x,\\
   \D-\I r_t=r_{xx}+(qr^2)_x.\\
   \end{array}\label{eq:EQs}
\end{equation}

For simplicity, for a function $f(x,t)$, denote $\bar
f(x,t)=f(-x,t)$. Note that $\D\frac{\partial}{\partial x}\bar
f(x,t)=-f_x(-x,t)$. Here $f_x$ refers to the partial derivative of
$f$ with respect to the first variable.

With these notations, we impose a relation $r=-\varepsilon\bar q^*$
where $\varepsilon=\pm 1$. Then the system (\ref{eq:EQs}) is reduced
to one equation --- the nonlocal derivative nonlinear Schr\"odinger
equation (\ref{eq:DNLS}).

With the reduction $r=-\varepsilon\bar q^*$, the coefficients of the
Lax pair satisfy
\begin{equation}
   \begin{array}{l}
   \D JU(\lambda)J^{-1}=U(-\lambda),\quad
   JV(\lambda)J^{-1}=V(-\lambda),\\
   \D KU(\lambda)K^{-1}=-(\bar U(-\lambda^*))^*,\quad
   KV(\lambda)K^{-1}=(\bar V(-\lambda^*))^*
   \end{array}\label{eq:sym}
\end{equation}
where $\D J=\left(\begin{array}{cc}1&0\\0&-1\end{array}\right)$, $\D
K=\left(\begin{array}{cc}0&-\varepsilon\\1&0\end{array}\right)$.
Here $U^*$ is the complex conjugation of the matrix $U$ without
transpose. We will use $U^\dagger=(U^*)^T$ in this paper.

Considering these symmetries of the Lax pair, we have immediately
the symmetries of the solutions of the Lax pair as follows.

\begin{lemma}\label{lemma:sym}
If $\Phi$ is a solution of the Lax pair (\ref{eq:LP}) with
$\lambda=\mu$, then $J\Phi$, $JK\bar\Phi^*$ and $K\bar\Phi^*$ are
solutions of (\ref{eq:LP}) with $\lambda=-\mu$, $\lambda=\mu^*$ and
$\lambda=-\mu^*$ respectively. Equivalently, if
$\left(\begin{array}{c}\xi\\\eta\end{array}\right)$ is a solution of
(\ref{eq:LP}) with $\lambda=\mu$, then
$\left(\begin{array}{c}\xi\\-\eta\end{array}\right)$,
$\left(\begin{array}{c}\varepsilon\bar\eta^*\\\bar\xi^*\end{array}\right)$
and
$\left(\begin{array}{c}-\varepsilon\bar\eta^*\\\bar\xi^*\end{array}\right)$
are solutions of (\ref{eq:LP}) with $\lambda=-\mu$, $\lambda=\mu^*$
and $\lambda=-\mu^*$ respectively.
\end{lemma}

\section{Darboux transformation of degree one}\label{sect:1}

\subsection{Darboux transformation for unreduced system}

We do not consider the reduction $r=-\varepsilon\bar q^*$
temporarily. Like that for the derivative nonlinear Schr\"odinger
equation,\cite{bib:Imai,bib:Steudel,bib:Zthesis} a Darboux
transformation of degree one can be constructed as follows.

\begin{lemma}\label{lemma:dt1}
Suppose $G(x,t,\lambda)=R(x,t)(\lambda-S(x,t))$ is a Darboux matrix
for (\ref{eq:LP}), which transforms $\D U=\lambda^2J+\lambda
P=\left(\begin{array}{cc}\lambda^2&\lambda q\\\lambda
r&-\lambda^2\end{array}\right)$ to $\widetilde
U=\lambda^2J+\lambda\widetilde P
=\left(\begin{array}{cc}\lambda^2&\lambda\widetilde q\\\lambda
\widetilde r&-\lambda^2\end{array}\right)$ and transforms $V$ in
(\ref{eq:LPcoeff}) to $\widetilde V$ which has the same form as $V$ where
$(q,r)$ are replaced by $(\widetilde q,\widetilde r)$, then $R$ is a diagonal matrix
and $RS$ is a constant matrix. Moreover, the transformation of $P$
is
\begin{equation}
   \widetilde P=RPR^{-1}+[J,RS]R^{-1}.\label{eq:qrmx}
\end{equation}
\end{lemma}

\begin{demo}
For the $x$-part, the condition $GU+G_x=\widetilde UG$, which means
that $G$ is a Darboux matrix, is
\begin{equation}
   R(\lambda-S)(\lambda^2J+\lambda P)+\lambda R_x-(RS)_x
   =(\lambda^2J+\lambda\widetilde P)R(\lambda-S).
\end{equation}
Compare the coefficients of the powers of $\lambda$. The coefficient
of $\lambda^3$ implies that $R$ is a diagonal matrix. The term
without $\lambda$ implies that $RS$ is independent of $x$. The
coefficient of $\lambda^2$ gives the transformation (\ref{eq:qrmx}).
On the other hand, $RS$ is also independent of $t$ by considering
the $t$ equation. The lemma is proved.
\end{demo}

By Lemma~\ref{lemma:sym}, if
$\D\left(\begin{array}{c}\xi\\\eta\end{array}\right)$ is a solution
of (\ref{lemma:sym}) with $\lambda=\mu$, then
$\D\left(\begin{array}{c}\xi\\-\eta\end{array}\right)$ is a solution
of (\ref{lemma:sym}) with $\lambda=-\mu$. Following the idea of
Gu\cite{bib:Gu,bib:GHZbook}, let
\begin{equation}
   \begin{array}{l}
   \D\Lambda=\left(\begin{array}{cc}\mu\\&-\mu\end{array}\right),\quad
   H=\left(\begin{array}{cc}\xi&\xi\\\eta&-\eta\end{array}\right),
   \end{array}\label{eq:LambdaH}
\end{equation}
then
\begin{equation}
   S\overset\triangle=H\Lambda H^{-1}
   =\mu\left(\begin{array}{cc}0&1/\sigma\\\sigma&0\end{array}\right)
   \label{eq:S}
\end{equation}
gives a Darboux matrix $G(\lambda)=R(\lambda I-S)$ where
$\sigma=\eta/\xi$, and $R$ is a suitable invertible matrix.

To make $R$ diagonal and $RS$ constant, choose
\begin{equation}
   R=\left(\begin{array}{cc}\sigma&0\\0&\D1/\sigma\end{array}\right),
   \label{eq:R}
\end{equation}
then $\D RS=\mu\left(\begin{array}{cc}0&1\\1&0\end{array}\right)$.
Then the Darboux matrix is
\begin{equation}
   \D G(\lambda)=R(\lambda-S)
   =\left(\begin{array}{cc}\sigma\lambda&-\mu\\-\mu&\D\lambda/\sigma
   \end{array}\right)
   \label{eq:G}
\end{equation}
and the transformation (\ref{eq:qrmx}) becomes
\begin{equation}
   \widetilde q=\sigma^2q+2\mu\sigma,\quad
   \widetilde r=\frac{r}{\sigma^2}-\frac{2\mu}{\sigma}.
   \label{eq:qr}
\end{equation}

\subsection{Darboux transformation for nonlocal derivative nonlinear
Schr\"o\-din\-ger equation}

Now we consider the reduction $r=-\varepsilon\bar q^*$.
Lemma~\ref{lemma:sym} implies that $-\mu$, $\mu^*$, $-\mu^*$ are all
eigenvalues if $\mu$ is. However, a Darboux transformation of degree
one only allows two eigenvalues. Hence $\mu$ should be chosen as
real or purely imaginary here. If $r=-\varepsilon\bar q^*$ and
$\widetilde r=-\varepsilon\bar{\widetilde q}^*$ hold in
(\ref{eq:qr}), then $\sigma\bar\sigma^*=\varepsilon\mu^*/\mu=\pm 1$.
The fact $\D\sigma\bar\sigma^*|_{(0,0)}=|\sigma(0,0)|^2\ge 0$
demands that $\sigma\bar\sigma^*=1$. Hence $\mu$ should be real if
$\varepsilon=1$ and purely imaginary if $\varepsilon=-1$.

From (\ref{eq:LP}) for $\lambda=\mu$, we have
\begin{equation}
   \begin{array}{l}
   \sigma_x=\mu r-2\mu^2\sigma-\mu q\sigma^2,\\
   \sigma_t=\I\mu(-2\mu^2r+r_x+qr^2)+2\I\mu^2(2\mu^2-qr)\sigma
   +\I\mu(2\mu^2q+q_x-q^2r)\sigma^2.
   \end{array}\label{eq:sigma}
\end{equation}
Hence
\begin{equation}
   \begin{array}{l}
   (\sigma\bar\sigma^*-1)_x=-\mu(r\bar\sigma^*+q\sigma)(\sigma\bar\sigma^*-1),\\\
   (\sigma\bar\sigma^*-1)_t=-\I\mu\left\{(-2\mu^2r+r_x+qr^2)\bar\sigma^*
   -(2\mu^2q+q_x-q^2r)\sigma\right\}(\sigma\bar\sigma^*-1),
   \end{array}
\end{equation}
when $\mu^*=\varepsilon\mu$. This guarantees that
$\sigma\bar\sigma^*=1$ holds identically if it holds at a specific
point $(x_0,t_0)$, say $(0,0)$.

In conclusion, the following theorem holds.

\begin{theorem}\label{thm1}
Suppose $q$ is a solution of (\ref{eq:DNLS}). Let $\mu$ be a nonzero
constant, which is real when $\varepsilon=1$ and purely imaginary
when $\varepsilon=-1$. Let $(\xi,\eta)^T$ be a solution of
(\ref{eq:LP}) with $\lambda=\mu$ such that $\sigma\bar\sigma^*=1$ at
the origin $(0,0)$ where $\sigma=\eta/\xi$. Then
\begin{equation}
   \widetilde q=\sigma^2q+2\mu\sigma \label{eq:q}
\end{equation}
is a new solution of (\ref{eq:DNLS}). The corresponding Darboux matrix is given
by (\ref{eq:G}).
\end{theorem}

\section{Darboux transformation of degree two}\label{sect:2}

\subsection{Darboux transformation for unreduced system}

Like (\ref{eq:LambdaH}), (\ref{eq:S}) and (\ref{eq:R}), take
\begin{equation}
   \begin{array}{l}
   \D\Lambda_\alpha
   =\left(\begin{array}{cc}\mu_\alpha\\&-\mu_\alpha\end{array}\right),\quad
   H_\alpha=\left(\begin{array}{cc}\xi_\alpha&\xi_\alpha\\
   \eta_\alpha&-\eta_\alpha\end{array}\right),\quad
   \sigma_\alpha=\frac{\eta_\alpha}{\xi_\alpha},\\
   \D S_\alpha=H_\alpha\Lambda_\alpha H_\alpha^{-1}
   =\mu\left(\begin{array}{cc}0&1/\sigma_\alpha\\
   \sigma_\alpha&0\end{array}\right),\quad
   R_\alpha=\left(\begin{array}{cc}\sigma_\alpha&0\\
   0&1/\sigma_\alpha\end{array}\right)\quad(\alpha=1,2).
   \end{array}\label{eq:LambdaH2}
\end{equation}
The Darboux matrix with respect to $(\Lambda_1,H_1)$ is given by
(\ref{eq:G}), i.e. $\D G_1(\lambda)=\left(\begin{array}{cc}\sigma_1 \lambda&-\mu_1\\
-\mu_1&\D\frac\lambda{\sigma_1}\end{array}\right)$.

After the action of $G_1(\lambda)$, the two columns of $H_2$ are
transformed to
\begin{equation}
   \begin{array}{l}
   \D
   G_1(\mu_2)\left(\begin{array}{cc}\xi_2\\\eta_2\end{array}\right)
   =\xi_2\left(\begin{array}{c}\mu_2\sigma_1-\mu_1\sigma_2\\
   \D-\mu_1+\frac{\mu_2\sigma_2}{\sigma_1}\end{array}\right)
   \overset\triangle=\left(\begin{array}{c}\widetilde\xi_2\\
   \widetilde\eta_2\end{array}\right),\\
   G_1(-\mu_2)\left(\begin{array}{cc}\xi_2\\-\eta_2\end{array}\right)
   =\xi_2\left(\begin{array}{c}-\mu_2\sigma_1+\mu_1\sigma_2\\
   \D-\mu_1+\frac{\mu_2\sigma_2}{\sigma_1}\end{array}\right)
   =-\left(\begin{array}{c}\widetilde\xi_2\\-\widetilde\eta_2\end{array}\right),
   \end{array}
\end{equation}
which are solutions of (\ref{eq:LP}) with potentials $\widetilde
q,\widetilde r$ where the eigenvalues are taken as $\lambda=\mu_2$
and $\lambda=-\mu_2$ respectively. Define
\begin{equation}
   \widetilde\sigma_2\overset\triangle=\frac{\widetilde\eta_2}{\widetilde\xi_2}
   =\frac1{\sigma_1}\frac{\mu_1\sigma_1-\mu_2\sigma_2}{\mu_1\sigma_2-\mu_2\sigma_1},
\end{equation}
then the Darboux matrix given by (\ref{eq:G}) for $(\widetilde
q,\widetilde r)$ is $\D \widetilde
G_2(\lambda)=\left(\begin{array}{cc}
\widetilde\sigma_2 \lambda&-\mu_2\\
-\mu_2&\D\frac\lambda{\widetilde\sigma_2}\end{array}\right)$. The
Darboux matrix of degree two for $(q,r)$ is
\begin{equation}
   \begin{array}{l}
   \D G(\lambda)=\widetilde
   G_2(\lambda)G_1(\lambda)
   =\left(\begin{array}{cc}\D\sigma_1\widetilde\sigma_2\lambda^2+\mu_1\mu_2
   &\D-\Big(\mu_1\widetilde\sigma_2+\frac{\mu_2}{\sigma_1}\Big)\lambda\\
   \D-\Big(\mu_2\sigma_1+\frac{\mu_1}{\widetilde\sigma_2}\Big)\lambda
   &\D\frac1{\sigma_1\widetilde\sigma_2}\lambda^2+\mu_1\mu_2\end{array}\right)\\
   \D=\left(\begin{array}{cc}
   \D\frac{\mu_1\sigma_1-\mu_2\sigma_2}{\mu_1\sigma_2-\mu_2\sigma_1}\lambda^2+\mu_1\mu_2
   &\D\frac{\mu_2^2-\mu_1^2}{\mu_1\sigma_2-\mu_2\sigma_1}\lambda\\
   \D\frac{(\mu_2^2-\mu_1^2)\sigma_1\sigma_2}{\mu_1\sigma_1-\mu_2\sigma_2}\lambda
   &\D\frac{\mu_1\sigma_2-\mu_2\sigma_1}{\mu_1\sigma_1-\mu_2\sigma_2}\lambda^2+\mu_1\mu_2
   \end{array}\right).
   \end{array}\label{eq:G2}
\end{equation}
By (\ref{eq:qr}), the solution of (\ref{eq:DNLS}) derived from this
$G(\lambda)$ is
\begin{equation}
   \begin{array}{l}
   \D\widetilde{\widetilde q}=\widetilde\sigma_2^2(\sigma_1^2q+2\mu_1\sigma_1)
   +2\mu_2\widetilde\sigma_2
   =\Big(\frac{\mu_1\sigma_1-\mu_2\sigma_2}{\mu_1\sigma_2-\mu_2\sigma_1}\Big)^2
   \Big(q+\frac{2(\mu_1^2-\mu_2^2)}{\mu_1\sigma_1-\mu_2\sigma_2}\Big),\\
   \D\widetilde{\widetilde r}=\frac1{\widetilde\sigma_2^2}\Big(\frac{r}{\sigma_1^2}
   -\frac{2\mu_1}{\sigma_1}\Big)-\frac{2\mu_2}{\widetilde\sigma_2}
   =\Big(\frac{\mu_1\sigma_2-\mu_2\sigma_1}{\mu_1\sigma_1-\mu_2\sigma_2}\Big)^2
   \Big(r-\frac{2(\mu_1^2-\mu_2^2)\sigma_1\sigma_2}{\mu_1\sigma_2-\mu_2\sigma_1}\Big).
   \end{array}\label{eq:qqrr}
\end{equation}
This result is similar to that in \cite{bib:Zhang} for usual
Kaup-Newell system.

\subsection{Darboux transformation for nonlocal derivative nonlinear
Schr\"o\-din\-ger equation}

According to Lemma~\ref{lemma:sym}, all the eigenvalues
$\mu,-\mu,\mu^*,-\mu^*$ should be considered in constructing Darboux
matrix when $\mu^2$ is not real. Therefore, a Darboux matrix of
degree two is necessary in this case.

Following Lemma~\ref{lemma:sym}, those in (\ref{eq:LambdaH2}) are
now
\begin{equation}
   \begin{array}{l}
   \mu_1=\mu,\quad\mu_2=\mu^*,\quad
   \xi_1=\xi,\quad \eta_1=\eta,\quad \xi_2=\varepsilon\bar\eta^*,\quad
   \eta_2=\bar\xi^*,\\
   \sigma_1=\sigma,\quad \sigma_2=\varepsilon/\bar\sigma^*.
   \end{array}
\end{equation}
The Darboux matrix is given by (\ref{eq:G2}), that is
\begin{equation}
   \begin{array}{l}
   \D G(\lambda)=\left(\begin{array}{cc}
   \D-\frac{\mu^*-\varepsilon\mu\sigma\bar\sigma^*}
   {\mu-\varepsilon\mu^*\sigma\bar\sigma^*}\lambda^2+|\mu|^2
   &\D\frac{\varepsilon(\mu^{*2}-\mu^2)\bar\sigma^*}{\mu-\varepsilon\mu^*\sigma\bar\sigma^*}
   \lambda\\
   \D-\frac{(\mu^{*2}-\mu^2)\sigma}
   {\mu^*-\varepsilon\mu\sigma\bar\sigma^*}\lambda
   &\D-\frac{\mu-\varepsilon\mu^*\sigma\bar\sigma^*}
   {\mu^*-\varepsilon\mu\sigma\bar\sigma^*}\lambda^2+|\mu|^2
   \end{array}\right).
   \end{array}\label{eq:DNLSG2}
\end{equation}
It can be checked that $G(\lambda)$ satisfies the reductions
$G(-\lambda)=J^{-1} G(\lambda)J$, $(\bar
G(-\lambda^*))^*=K^{-1}G(\lambda)K$, which are compatible with
(\ref{eq:sym}).

The new solution is given by (\ref{eq:qqrr}). That is

\begin{theorem}\label{thm2}
Suppose $q$ is a solution of (\ref{eq:DNLS}). Let $\mu$ be a nonzero
complex constant which is neither real nor purely imaginary. Let
$(\xi,\eta)^T$ be a solution of (\ref{eq:LP}) with $\lambda=\mu$.
Then
\begin{equation}
   \begin{array}{l}
   \D\widetilde q
   =\Big(\frac{\mu^*-\varepsilon\mu\sigma\bar\sigma^*}
   {\mu-\varepsilon\mu^*\sigma\bar\sigma^*}\Big)^2
   \Big(q-\frac{2(\mu^2-\mu^{*2})}{\mu^*-\varepsilon\mu\sigma\bar\sigma^*}
   \varepsilon\bar\sigma^*\Big)
   \end{array}\label{eq:DNLSq2}
\end{equation}
is a new solution of (\ref{eq:DNLS}) where $\sigma=\eta/\xi$. The
corresponding Darboux matrix is given by (\ref{eq:DNLSG2}).
\end{theorem}

\section{Darboux transformation of degree $2n$}\label{sect:2n}

The Darboux transformation of degree two is the Darboux
transformation of lowest degree that keeps all the reductions of
nonlocal derivative nonlinear Schr\"odinger equation if $\mu^2$ is
not real. The composition of these Darboux transformations leads to
a Darboux transformation of higher degree. However, this can be
constructed equivalently and more compactly following the idea of
Zakharov and Mihailov\cite{bib:Zak,bib:GHZbook}.

\begin{theorem}\label{thm2n}
Take $n$ complex numbers $\mu_j$ so that $\mu_j^2$'s are not real
and take the solution $(\xi_j,\eta_j)^T$ of the Lax pair
(\ref{eq:LP}) with $\lambda=\mu_j$ $(j=1,2,\cdots,n)$. Let
$\lambda_{2j-1}=\mu_j$, $\lambda_{2j}=-\mu_j$,
$h_{2j-1}=(\xi_j,\eta_j)^T$, $h_{2j}=(\xi_j,-\eta_j)^T$
$(j=1,2,\cdots,n)$. Let
\begin{equation}
   \Gamma_{\alpha\beta}
   =\frac{\bar h_\alpha^\dagger Lh_\beta}{\lambda_\alpha^*+\lambda_\beta}\quad
   (\alpha,\beta=1,\cdots,2n),
   \label{eq:Gamma}
\end{equation}
\begin{equation}
   G(\lambda)=\prod_{\gamma=1}^{2n}(\lambda+\lambda_\gamma^*)\cdot F^{-1}
   \left(I-\sum_{\alpha,\beta=1}^{2n}\frac{(\Gamma^{-1})_{\alpha\beta}
   h_\alpha\bar h_\beta^\dagger L}{\lambda+\lambda_\beta^*}\right)
   \label{eq:dthigh}
\end{equation}
where $\D
L=\left(\begin{array}{cc}\varepsilon\\&1\end{array}\right)$ and
\begin{equation}
   F=I-\sum_{\alpha,\beta=1}^{2n}\frac{(\Gamma^{-1})_{\alpha\beta}
   h_\alpha\bar h_\beta^\dagger L}{\lambda_\beta^*}\label{eq:Fdef}
\end{equation}
is a diagonal matrix. Then $G(\lambda)$ is a Darboux matrix for the
Lax pair (\ref{eq:LP}), and the transformation of $q$ is given by
\begin{equation}
   \widetilde q=\frac{\D I-\sum_{\alpha,\beta=1}^{2n}\frac{(\Gamma^{-1})_{\alpha\beta}
   \eta_\alpha\bar\eta_\beta^*}{\lambda_\beta^*}}
   {\D I-\varepsilon\sum_{\alpha,\beta=1}^{2n}\frac{(\Gamma^{-1})_{\alpha\beta}
   \xi_\alpha\bar\xi_\beta^*}{\lambda_\beta^*}}
   \Big(q+2\sum_{\alpha,\beta=1}^{2n}
   (\Gamma^{-1})_{\alpha\beta}\xi_\alpha\bar\eta_\beta^*\Big).\label{eq:q2n}
\end{equation}
\end{theorem}

\begin{remark}
(i) $G(\lambda)$ is a polynomial of $\lambda$ of degree $2n$.
Moreover, $\D G(0)=(-1)^n\prod_{l=1}^{n}\mu_l^{*2}I$ is a constant
scalar matrix. This is necessary since $G(0)$ of a Darboux matrix
(\ref{eq:DNLSG2}) of degree two is constant, so is their product.

(ii) (\ref{eq:dthigh}) gives the same Darboux matrix as
(\ref{eq:DNLSG2}) up to a constant scalar multiplier when $n=1$.
\end{remark}

\demo We have
\begin{equation}
   \begin{array}{l}
   \D\left(I-\sum_{\alpha,\beta=1}^{2n}
   \frac{(\Gamma^{-1})_{\alpha\beta}h_\alpha\bar h_\beta^\dagger L}
   {\lambda+\lambda_\beta^*}\right)\Big|_{\lambda=\lambda_\gamma}h_\gamma
   =h_\gamma-\sum_{\alpha,\beta=1}^{2n}
   \frac{(\Gamma^{-1})_{\alpha\beta}h_\alpha\bar h_\beta^\dagger Lh_\gamma}
   {\lambda_\gamma+\lambda_\beta^*}=0,\\
   \D(\lambda+\lambda_\gamma^*)\left(I-\sum_{\alpha,\beta=1}^{2n}
   \frac{(\Gamma^{-1})_{\alpha\beta}h_\alpha\bar h_\beta^\dagger L}
   {\lambda+\lambda_\beta^*}\right)
   \Big|_{\lambda=-\lambda_\gamma^*}K\bar
   h_\gamma^*=0
   \end{array}
\end{equation}
since $\bar h_\gamma^\dagger LK\bar h_\gamma^*=0$. Hence
\begin{equation}
   G(\lambda_\gamma)h_\gamma=0,\quad G(-\lambda_\gamma^*)K\bar h_\gamma^*=0\quad
   (\gamma=1,2.\cdots,2n).
\end{equation}
$h_\gamma$ is a solution of the Lax pair (\ref{eq:LP}) with
$\lambda=\lambda_\gamma$, Lemma~\ref{lemma:sym} implies that $K\bar
h_\gamma^*$ is a solution of (\ref{eq:LP}) with
$\lambda=-\lambda_\gamma^*$. By the standard construction of Darboux
transformation\cite{bib:GHZbook}, $G(\lambda)$ is a composition of
$2n$ Darboux matrices in the form (\ref{eq:G}). From the choice of
$\lambda_j$'s and $h_j$'s, $G(\lambda)$ is a composition of $n$
Darboux matrices in the form (\ref{eq:G2}). That is,
\begin{equation}
   G(\lambda)=(-1)^nG_n(\lambda)\cdots G_1(\lambda)\prod_{l=1}^n\frac{\mu^*_l}{\mu_l}
   \label{eq:Gdecomp}
\end{equation}
where each $G_l(\lambda)$ $(l=1,\cdots,n)$ is a Darboux matrix in
the form (\ref{eq:G2}). Hence $G(\lambda)$ keeps the reduction
$r=-\varepsilon\bar q^*$, i.e. it is a Darboux matrix for the
nonlocal derivative nonlinear Schr\"odinger equation.

Write
\begin{equation}
   G(\lambda)=G_0\lambda^{2n}+G_1\lambda^{2n-1}+\cdots+G_{2n},
\end{equation}
then the coefficient of $\lambda^{2n+1}$ in
$G(\lambda)U(\lambda)+G_x(\lambda)=\widetilde U(\lambda)G(\lambda)$
gives the transformation
\begin{equation}
   \widetilde P=G_0PG_0^{-1}-[J,G_1]G_0^{-1}.\label{eq:P2n}
\end{equation}
Expanding (\ref{eq:dthigh}) as a polynomial of $\lambda$ and using
$\D\sum_{\gamma=1}^{2n}\lambda_\gamma^*=0$, we get
\begin{equation}
   G_0=F^{-1},\quad
   G_1=-F^{-1}\sum_{\alpha,\beta=1}^{2n}(\Gamma^{-1})_{\alpha\beta}
   h_\alpha\bar h_\beta^*L.
\end{equation}
Since each $G_l(\lambda)$ in (\ref{eq:Gdecomp}) is in the form
(\ref{eq:G2}), $F=G_0^{-1}$ must be diagonal. (This can also be
verified algebraically from the expression (\ref{eq:Fdef}) by
exchanging each pair of $(2j-1,2j)$ $(j=1,\cdots,n)$ in the
subscripts.) The $(1,2)$ entry of (\ref{eq:P2n}) gives
(\ref{eq:q2n}). The theorem is proved.

\section{Examples}\label{sect:eg}

When $q=r=0$, the Lax pair (\ref{eq:LP}) becomes
\begin{equation}
   \Phi_x=\left(\begin{array}{cc}\lambda^2&0\\
   0&-\lambda^2\end{array}\right)\Phi,\quad
   \Phi_t=\left(\begin{array}{cc}-2\I\lambda^4&0\\
   0&2\I\lambda^4\end{array}\right)\Phi.\label{eq:egLP}
\end{equation}
The solution of this system for $\lambda=\mu$ is
$\D\left(\begin{array}{c}\xi\\\eta\end{array}\right)
=\left(\begin{array}{c}a\E{\theta}\\
b\E{-\theta}\end{array}\right)$ where $\theta=\mu^2x-2\I\mu^4t$ and
$a,b$ are complex constants. Then
\begin{equation}
   \sigma=\frac{\eta}{\xi}=c\E{-2\theta} \label{eq:sigmaeg}
\end{equation}
where $c=b/a$ is a complex constant.

The behavior of the solutions depends on $4\arg\mu/\pi$. When
$4\arg\mu/\pi$ is an even integer, then $\mu^2$ is real. A Darboux
matrix of degree one can be constructed as in Theorem~\ref{thm1}.
When $4\arg\mu/\pi$ is not an integer, a Darboux matrix of degree
two should be used as in Theorem~\ref{thm2}. However, the new
solutions constructed from zero solution is unbounded in the first
case and have singularities in the second case. We are not
interested in these examples. Thus we only consider the case when
$4\arg\mu/\pi$ is an odd integer. Without loss of generality, take
$\arg\mu_j=\pi/4$ $(j=1,\cdots,n)$ for a Darboux transformation of
degree $2n$.

\begin{figure}\begin{center}
\scalebox{1.3}{\includegraphics[250,200]{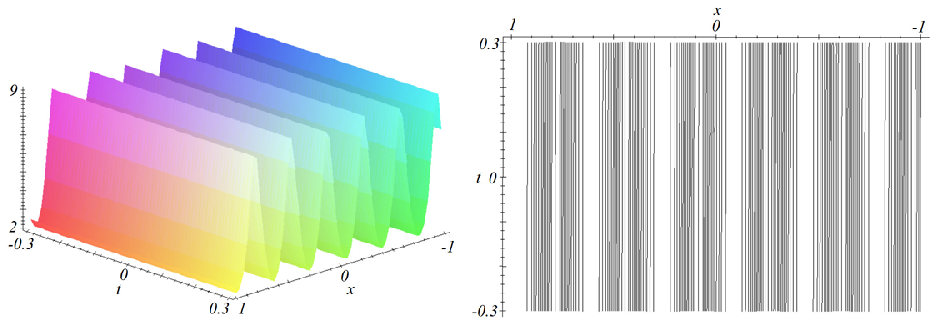}}
\caption{$|\widetilde q|$: $\widetilde q$ is given by a Darboux
transformation of degree $2$}\label{fig1} \end{center}\end{figure}

First consider a Darboux transformation of degree $2$ with
$\mu=a(1+\I)$ where $a$ is a positive constant. By
(\ref{eq:DNLSq2}),
\begin{equation}
   \widetilde q=-4(1-\I)\varepsilon ac^*
   \frac{\E{-4\I a^2x}(1-\I\varepsilon|c|^2\E{-8\I a^2x})}
   {(1+\I\varepsilon|c|^2\E{-8\I a^2x})^2}\E{16\I a^4t}.
   \label{eq:qeg1}
\end{equation}
This solution is global if $|c|\ne 1$. It is periodic in both $x$
and $t$, and $|\widetilde q|$ is a function of $x$ only.
Figure~\ref{fig1} shows the norm of the solution with
$\varepsilon=1$, $\mu=1.5(1+\I)$, $c=0.5$. Hereafter, the figure on
the right shows the contour plot of the one on the left.

\begin{figure}\begin{center}
\scalebox{1.3}{\includegraphics[250,200]{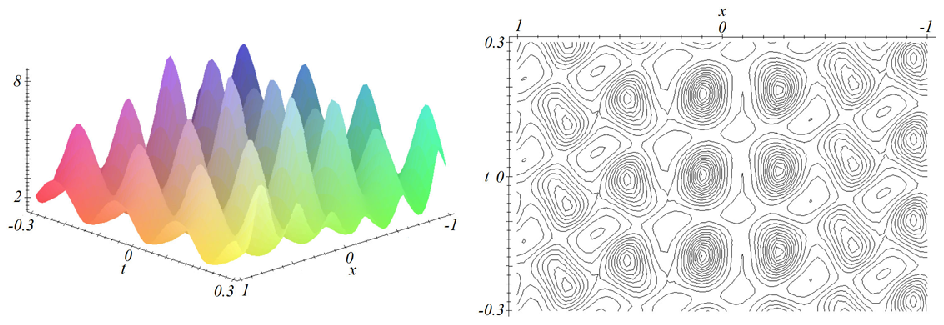}}
\caption{$|\widetilde q|$: $\widetilde q$ is given by a Darboux
transformation of degree $4$}\label{fig2}
\end{center}\end{figure}

\begin{figure}\begin{center}
\scalebox{1.3}{\includegraphics[250,200]{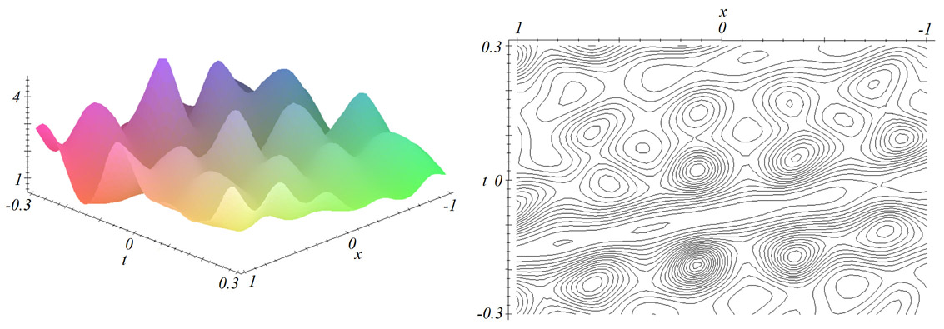}}
\caption{$|\widetilde q|$: $\widetilde q$ is given by a Darboux
transformation of degree $8$}\label{fig4} \end{center}\end{figure}

By using Theorem~\ref{thm2n}, the figures for the norm of the
solutions given by Darboux transformations of degree $4$ and $8$ are
plotted in Figure~\ref{fig2} ($\varepsilon=1$, $a_1=1.5$, $a_2=1.3$,
$c_1=0.05$, $c_2=-0.02$) and Figure~\ref{fig4} ($\varepsilon=1$,
$a_1=1.5$, $a_2=1.3$, $a_3=1.1$, $a_4=0.9$, $c_1=0.001$,
$c_2=-0.002$, $c_3=0.001\I$, $c_4=-0.001$).

\section{Globalness of the solutions}\label{sect:global}

Although the solutions (\ref{eq:qeg1}) given by Darboux
transformation of degree two are always global when $|c|\ne 1$,
those given by Darboux transformation of higher degree may not.
However, we can prove that the solutions have no singularities when
all $|c_j|$'s are small enough.

Before proving that theorem, we need the follow algebraic lemma.

\begin{lemma}\label{lemma:alg}
Let $\lambda_1,\cdots,\lambda_m,\omega_1,\cdots,\omega_m$ be nonzero
complex constants such that $\omega_j+\lambda_k\ne 0$
$(j,k=1,\cdots,m)$, and $\D C=\Big(\frac
1{\omega_j+\lambda_k}\Big)_{1\le j,k\le m}$. Then
\begin{equation}
   1-\sum_{1\le j,k\le m}(C^{-1})_{jk}\frac 1{\omega_k}
   =\prod_{l=1}^m\Big(-\frac{\lambda_l}{\omega_l}\Big).
\end{equation}
\end{lemma}

\demo By the formula for computing the determinant of a block
matrix,
\begin{equation}
   \begin{array}{l}
   \D1-\sum_{1\le j,k\le m}(C^{-1})_{jk}\frac 1{\omega_k}
   \D=\frac 1{\det C}\left|\begin{array}{cccc}
   \D\frac1{\omega_1+\lambda_1}&\cdots&\D\frac1{\omega_1+\lambda_m}&\D\frac 1{\omega_1}\\
   \vdots&&\vdots&\vdots\\
   \D\frac1{\omega_m+\lambda_1}&\cdots&\D\frac1{\omega_m+\lambda_m}&\D\frac 1{\omega_{m}}\\
   1&\cdots&1&1\\
   \end{array}\right|
   \D\overset\triangle=\frac{\Delta}{\det C}.
   \end{array}
\end{equation}
Clearly,
\begin{equation}
   \Delta=\frac{\D T(\lambda_1,\cdots,\lambda_{m},\omega_1,\cdots,\omega_{m})}
   {\D\prod_{l=1}^{m}\omega_l\prod_{1\le j,k\le m}(\omega_j+\lambda_k)}
\end{equation}
where $T$ is a polynomial of degree $m^2$. Note that $T=0$ when
$\omega_j=\omega_k$ or $\lambda_j=\lambda_k$ for certain $j\ne k$
with $1\le j,k\le m$, or when $\lambda_l=0$ for certain $l$ with
$1\le l\le m$. Considering the degree of the polynomial, we have
\begin{equation}
   T=\rho\prod_{l=1}^m\lambda_l\prod_{1\le j<k\le m}(\lambda_j-\lambda_k)(\omega_j-\omega_k)
\end{equation}
where $\rho$ is a constant. When $\lambda_1,\cdots,\lambda_m$ are
distinct,
\begin{equation}
   1=\lim_{\omega_1\to-\lambda_1,\cdots,\omega_m\to-\lambda_m}
   \Delta\prod_{j=1}^m(\omega_j+\lambda_j)
   =(-1)^m\rho
\end{equation}
which leads to $\rho=(-1)^m$. Likewise, we have
\begin{equation}
   \det C=\frac{\D\prod_{1\le j<k\le m}(\lambda_j-\lambda_k)(\omega_j-\omega_k)}
   {\D\prod_{1\le j,k\le
   m}(\omega_j+\lambda_k)},\label{eq:Cdet}
\end{equation}
which is the standard Cauchy determinant. Therefore,
\begin{equation}
   1-\sum_{1\le j,k\le m}(C^{-1})_{jk}\frac
   1{\omega_k}=\prod_{l=1}^m\Big(-\frac{\lambda_l}{\omega_l}\Big).
\end{equation}
The lemma is proved.

\begin{theorem}
Take the seed solution $q=0$. Suppose $\mu_j=a_j\E{\pi\I/4}$ where
$a_1,\cdots,a_n$ are distinct positive numbers. Then there is a
positive constant $\delta$ such that the solution (\ref{eq:q2n})
given by a Darboux transformation of degree $2n$ is globally defined
and bounded for $(x,t)\in\hr^2$ when $|c_j|<\delta$
$(j=1,\cdots,n)$.
\end{theorem}

\demo It is easy to check from (\ref{eq:Gamma}) and
(\ref{eq:dthigh}) that $G(\lambda)$ is invariant if each
$(\xi_j,\eta_j)$ is changed to $\rho_j(\xi_j,\eta_j)$ for any
nonzero constants $\rho_j$ $(j=1,\cdots,n)$. Hence $(\xi_j,\eta_j)$
can be replaced by $(1,\sigma_j)$.

It is only necessary to prove that $\Gamma$ in (\ref{eq:Gamma}) and
$F$ in (\ref{eq:Fdef}) are both invertible for all $(x,t)$ when
$|c_j|$ $(j=1,\cdots,n)$ are small enough.

As before, denote $h_{2j-1}=(1,\eta_j)^T$, $h_{2j}=(1,-\eta_j)^T$.
Since $\sigma_j=\E{-2\theta_j}=\E{-2\mu_j^2x+4\I\mu_k^4t}$,
$\sigma_k\bar\sigma_j^*=c_kc_j^*\E{-\I\phi_{jk}}$ where
$\phi_{jk}=2(a_j^2+a_k^2)x+4(a_j^4-a_k^4)t$ is real. Hence
\begin{equation}
   \begin{array}{l}
   \D\bar h_{2j-1}^\dagger Lh_{2k-1}=\bar h_{2j}^\dagger Lh_{2k}
   =\varepsilon+\sigma_k\bar\sigma_j^*
   =\varepsilon(1+\varepsilon c_kc_j^*\E{-\I\phi_{jk}}),\\
   \D\bar h_{2j-1}^\dagger Lh_{2k}=\bar h_{2j}^\dagger Lh_{2k-1}=
   \varepsilon-\sigma_k\bar\sigma_j^*
   =\varepsilon(1-\varepsilon c_kc_j^*\E{-\I\phi_{jk}}).
   \end{array}
\end{equation}
$\Gamma=(\Gamma_{\alpha\beta})_{2n\times 2n}$ in (\ref{eq:Gamma})
can be written as a block matrix
$\D\Gamma=(\widetilde\Gamma_{jk})_{n\times n}$ where
\begin{equation}
   \widetilde\Gamma_{jk}=\left(\begin{array}{cc}
   \D\frac{\bar h_{2j-1}^\dagger Lh_{2k-1}}{\mu_j^*+\mu_k}
   &\D\frac{\bar h_{2j-1}^\dagger Lh_{2k}}{\mu_j^*-\mu_k}\\
   \D\frac{\bar h_{2j}^\dagger Lh_{2k-1}}{-\mu_j^*+\mu_k}
   &\D\frac{\bar h_{2j}^\dagger Lh_{2k}}{-\mu_j^*-\mu_k}\end{array}\right)
   \!\!\!=\!\varepsilon\left(\begin{array}{cc}
   \D\frac{1+\varepsilon c_kc_j^*\E{-\I\phi_{jk}}}{\mu_j^*+\mu_k}
   &\D\frac{1-\varepsilon c_kc_j^*\E{-\I\phi_{jk}}}{\mu_j^*-\mu_k}\\
   \D\frac{1-\varepsilon c_kc_j^*\E{-\I\phi_{jk}}}{-\mu_j^*+\mu_k}
   &\D\frac{1+\varepsilon c_kc_j^*\E{-\I\phi_{jk}}}{-\mu_j^*-\mu_k}\\
   \end{array}\right).
\end{equation}
Hence $\Gamma$ converges to $\Gamma|_{c_1=\cdots=c_n=0}$ uniformly
for $(x,t)\in\hr^2$ when $c_1\to 0,\cdots,c_n\to 0$. Note that
\begin{equation}
   \Gamma|_{c_1=\cdots=c_n=0}
   =\varepsilon\Big(\frac 1{\lambda_j^*+\lambda_k}\Big)_{1\le j,k\le 2n}
   \label{eq:Gamma0}
\end{equation}
where $\lambda_{2j-1}=\mu_j$, $\lambda_{2j}=-\mu_j$
$(j=1,\cdots,n)$. Its determinant is
\begin{equation}
   \det\Gamma|_{c_1=\cdots=c_n=0}
   =\frac{\D\prod_{1\le \alpha<\beta\le 2n}|\lambda_\alpha-\lambda_\beta|^2}
   {\D\prod_{1\le\alpha,\beta\le 2n}(\lambda_\alpha^*+\lambda_\beta)}\ne 0
   \label{eq:Gamma0det}
\end{equation}
as (\ref{eq:Cdet}). Hence $\Gamma$ is invertible when $|c_j|$
$(j=1,\cdots,n)$ are small enough.

According to Lemma~\ref{lemma:alg} and (\ref{eq:Gamma0}),
\begin{equation}
   \begin{array}{l}
   \D F|_{c_1=\cdots=c_n=0}=I-\sum_{\alpha,\beta=1}^{2n}
   \frac{\D(\Gamma|_{c_1=\cdots=c_n=0}^{-1})_{\alpha\beta}
   \left(\begin{array}{cc}1&0\end{array}\right)
   \left(\begin{array}{c}1\\0\end{array}\right)
   \left(\begin{array}{cc}\varepsilon\\&1\end{array}\right)}{\lambda_\beta^*}\\
   \D=\left(\begin{array}{cc}1\\&1\end{array}\right)
   -\sum_{\alpha,\beta=1}^{2n}
   \frac{\D((\varepsilon\Gamma)|_{c_1=\cdots=c_n=0}^{-1})_{\alpha\beta}}{\lambda_\beta^*}
   \left(\begin{array}{cc}1\\&0\end{array}\right)\\
   \D=\left(\begin{array}{cc}\D\prod_{\gamma=1}^{2n}
   \frac{\lambda_\gamma}{\lambda_\gamma^*}&0\\0&1\end{array}\right)
   =\left(\begin{array}{cc}\D\prod_{l=1}^{n}\Big(\frac{\mu_l}{\mu_l^*}\Big)^2&0\\
   0&1\end{array}\right)=\left(\begin{array}{cc}(-1)^n&0\\0&1\end{array}\right)
   \end{array}\label{eq:F0}
\end{equation}
is invertible. (\ref{eq:Gamma0det}) implies that $\Gamma^{-1}$ is
bounded for all $(x,t)$ when $|c_j|$ $(j=1,\cdots,n)$ are small
enough. Hence $F$ converges to $F|_{c_1=\cdots=c_n=0}$ uniformly for
$(x,t)\in\hr^2$ when $c_1\to 0,\cdots,c_n\to 0$, which implies that
$F$ is invertible if $|c_j|$ $(j=1,\cdots,n)$ are small enough.

Finally, by the expression (\ref{eq:q2n}), boundedness of
$\widetilde q$ follows from (\ref{eq:Gamma0det}), (\ref{eq:F0}) and the
boundedness of $\sigma_j$'s. The theorem is proved.

\begin{remark}
By considering $\sigma^{-1}=\eta/\xi$ instead of $\sigma=\xi/\eta$,
it is easy to see that the above theorem is also true if all $|c_j|$
$(j=1,\cdots,n)$ are large enough.
\end{remark}

\begin{remark}
For the usual derivative nonlinear Schr\"odinger equation
(\ref{eq:DNLS0}) or the nonlocal nonlinear Schr\"odinger equation,
interesting global solutions can be derived from bounded exponential
seed solutions. However, there is no such kind of seed solution for
(\ref{eq:DNLS}). It is worth finding other interesting bounded seed
solutions for the nonlocal derivative nonlinear Schr\"odinger
equation (\ref{eq:DNLS}).

\end{remark}

\section*{Acknowledgements}

This work was supported by the Natural Science Foundation of
Shanghai (No.\ 16ZR\-1402600) and the Key Laboratory of Mathematics
for Nonlinear Sciences of Ministry of Education of China.

\end{document}